\documentclass[pdflatex,sn-mathphys-num]{sn-jnl}% Math and Physical Sciences Numbered Reference Style
\usepackage[utf8]{inputenc}
\usepackage[T1]{fontenc}
\usepackage{amssymb,amsmath,gensymb}
\usepackage[normalem]{ulem}
\usepackage{changes}
\usepackage{chngcntr}
\usepackage{epsfig}
\usepackage{amssymb}
\usepackage{soul}
\usepackage{color}
\usepackage{float}
\usepackage{textcomp}
\usepackage{amstext}
\usepackage{graphicx}
\usepackage{epstopdf}
\usepackage{gensymb}
\usepackage{graphics}\usepackage{subfigure}\usepackage{longtable}\usepackage{pstricks}\usepackage{dcolumn}\usepackage{bm}
\usepackage{siunitx}
\usepackage{wasysym}
\usepackage[font=small,labelfont=bf]{caption}

\theoremstyle{thmstyleone}%
\theoremstyle{thmstyletwo}%

\theoremstyle{thmstylethree}%

\begin{document}

%\title[Article Title]{Tunable Spin Dynamics in Entangled Muon–Fluorine Systems via RF Pulse Excitation}
\title[Vector spin polarization]{Vector spin polarization evolution determined in an entangled muon-fluorine system under pulsed excitation}


\author[1]{\fnm{Dipranjan} \sur{Chatterjee}}%%\\email{dipranjan.chatterjee@physics.ox.ac.uk}

\author[1]{\fnm{ Benjamin M.} \sur{Huddart}}%%\email{benjamin.huddart@physics.ox.ac.uk}

\author[1]{\fnm{Hank C. H. } \sur{Wu}}%%\email{benjamin.huddart@physics.ox.ac.uk}

\author[1]{\fnm{Dharmalingam} \sur{Prabhakaran}}%%\email{dharmalingam.prabhakaran@physics.ox.ac.uk}

\author[2]{\fnm{Alex} \sur{Louat}}

\author[2]{\fnm{Stephen P.} \sur{Cottrell}}

\author[1]{\fnm{Stephen} \sur{J. Blundell}}%%\email{stephen.blundell@physics.ox.ac.uk}

\affil[1]{Clarendon Laboratory, Department of Physics, University of Oxford, Parks Road, OX1 3PU, United Kingdom}

\affil[2]{STFC-ISIS Facility, Rutherford Appleton Laboratory, Harwell Campus, Chilton,
Oxfordshire, OX11 0QX, United Kingdom}


\abstract{A spin-polarized muon implanted into a fluoride forms a coupled F--$\mu$--F complex in which the muon spin and neighbouring fluorine nuclear spins become entangled. Here we apply radio-frequency (RF) excitation to this coupled system and 
use the three-dimensional distribution of emitted positrons to reconstruct the time-dependent evolution of the muon spin polarization.
%This time-dependent signal is a direct result of the interactions between the individual spins in the system and the RF pulses. 
This three-dimensional readout, using single spin detection, is not possible in a single NMR experiment and demonstrates significant advantages that are achieved by using RF muon techniques.  We demonstrate the application of this vector-readout method to the experimental observation of a muon spin echo signal that is controlled by the dipolar coupling to fluorine, as well as to a double resonance experiment, in which we use pulses tuned to separate frequencies to address both the muon and fluorine spins.
This targeted approach, in which selective RF pulses can control the muon spin and other spins to which it is coupled, provides a novel route for probing systems of entangled spins. 
}

\keywords{muon spin rotation, spin echo, entanglement, quantum control}



\maketitle

%\section* {Introduction}\label{sec1}
External control in local probe measurements introduces an additional dimension in exploring quantum phenomena such as coherence and entanglement, offering a deeper insight into the local environments within quantum materials. Techniques like nuclear magnetic resonance (NMR) have been instrumental in developing methods for quantum control and coherence studies, serving as foundational tools 
that inform the principles of quantum technology \cite{chuang1998experimental,jones2000nmr, knill2000theory,price1999construction,somaroo1999quantum,leskowitz2003three}. In conventional NMR the detection coil only picks up magnetization that is precessing transverse to the static field $B_0$ (conventionally aligned along the $z$ axis).  The components $M_x$ and $M_y$ produce an oscillating voltage in the detection coil, while the longitudinal component $M_z$ is invisible to the receiver \cite{Slitcher, Levitt}. To recover $M_z$, one must interrupt the experiment with separate pulse sequences of inversion recovery or saturation recovery and wait for relaxation, so that one cannot measure all three components of $\mathbf{M}$ at once.

In this paper, we demonstrate a method that allows us to monitor the complete vector polarization, while simultaneously permitting control and detection of local spin interactions  
as well as the transfer of coherence between local subsystems.  This is achieved using pulsed RF excitation in a muon spin relaxation ($\mu$SR) experiment \cite{Clayden2012}.  In this technique,
the readout signal is extracted from measuring positron counts, and hence it is possible to record the time dependence of the three-dimensional vector polarization of the muon even during the application of an RF pulse. This is not possible with nuclear polarization in NMR since the detector circuits in an NMR experiment would be saturated during excitation and so measurement is only possible after RF excitation has ceased.  
When implanted into a sample of an insulating fluoride, such as $\mathrm{LaF}_3$, a $\mu^+$ occupies an interstitial site between two fluorine ions and forms a linear F--$\mu$--F complex \cite{PhysRevBBrewerF-mu-F, PhysRevLett.99.267601, PhysRevLett.125.087201}.  The muon ($S=\frac{1}{2}$) is coupled to these $^{19}$F nuclear spins ($I=\frac{1}{2}$) via dipolar interactions. Previous RF-$\mu$SR studies in fluorides have followed two main approaches: one focused on analyzing effective spin echo responses with an emphasis on muon diffusion effects \cite{PhysRevLett.61.2890}, and another employing continuous-wave RF excitation to drive transitions in the F--$\mu$--F system \cite{PhysRevLettBillington}. In contrast, the present study utilizes pulsed RF Hahn echo measurements to achieve precise and selective inversion of spin state populations and coherent Rabi driving in the entangled F--$\mu$--F complex, monitored via the full three-dimensional readout of the muon polarization vector. These experimental results are supported by detailed computational modeling of the local spin interactions and dynamics.

\begin{figure}[t]
	\begin{centering}
		\includegraphics[width=1.0\columnwidth]{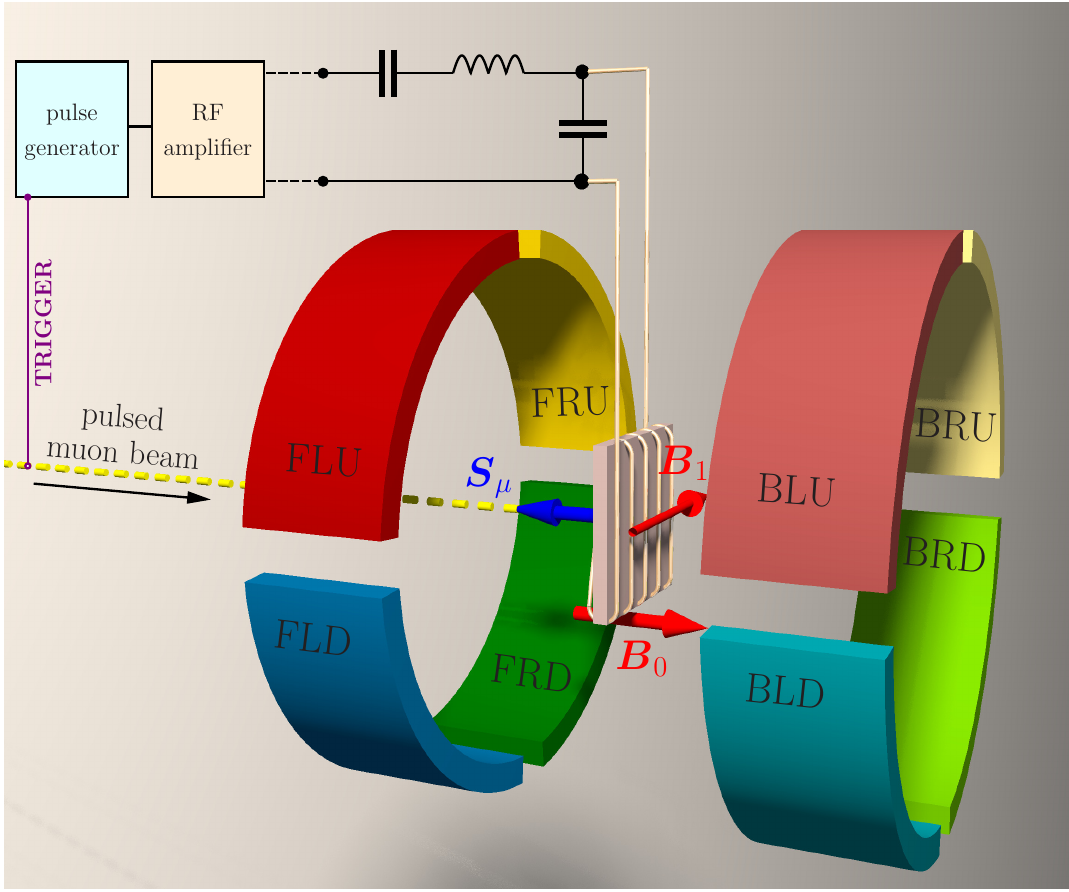} 
		\par\end{centering}
	\caption{Schematic of experiment.  The muon is implanted with its spin antiparallel to its momentum, with the pulse of the muon synchronised with the pulse generator that controls the radio-frequency (RF) excitation.  A steady magnetic field $B_0$ is applied longitudinally (aligned with the initial muon spin direction).  The RF amplifier and tuned circuit result in an oscillating field $B_1$ which is applied in a direction transverse to the initial muon spin direction.  The detectors are segmented into eight octants.  Their outputs can be combined in various ways in software, thereby realising a three-dimensional readout of the muon spin polarization as a function of time.
    The forward (F) detector is assembled using FLU+FRU+FLD+FRD; the left (L) detector is FLU+FRU+BLU+BRU; the down (D) detector by FLD+FRD+BLD+BRD, etc.}
	\label{fig:experiment}
\end{figure}

A schematic of the experimental arrangement is shown in Fig.~\ref{fig:experiment}.  The 96-detector array of the EMU spectrometer at the ISIS Pulsed Muon Source was used, grouped into eight segments of 12 detectors.  By adding these segments in various combinations, it is possible to obtain 
forward/backward, top/bottom and left/right pairs that sample the muon polarization projections $P_z$, $P_y$ and $P_x$, respectively. In this configuration the three non-degenerate F--$\mu$--F transition frequencies imprint distinct oscillatory beats on each detector pair, so that from a single pulse sequence we simultaneously obtain time-resolved ${P}_x(t)$, ${P}_y(t)$ and ${P}_z(t)$, providing the complete polarization-vector dynamics of the F--$\mu$--F complex in one measurement. 

\section* {RF amplitude ($B_1$) modulation with driving power}
In an RF experiment, a static longitudinal field $B_0$ is applied along the initial muon polarization direction and an oscillatory field $B_1 \cos (\omega_{\rm RF}t$) is applied in the transverse direction (see Fig.~\ref{fig:experiment}).  At resonance, the angular frequency of the oscillatory field is $\omega_{\rm RF}= \gamma B_0$. 
To perform a well-characterized Hahn-echo NMR experiment \cite{Hahn1950}, several experimental parameters must be carefully tuned, the most important being $B_\text{1}$. It fixes the flip angle $\theta=\gamma B_\text{1} t_\text{p}$ for a resonant RF pulse of width $t_\text{p}$, and therefore determines the pulse timing and excitation bandwidth.   It also sets the Rabi rate $\Omega_\text{R}=\gamma B_\text{1}$ and therefore
controls the change of populations of levels within the entangled F--$\mu$--F manifold.  The spatial uniformity of $B_1$, together with the resonator quality factor and the overall fidelity of the pulse, is what ultimately limits the accuracy of the flip angle and the quality of the echo. Hence, before performing any pulsed excitation, we first recorded continuous-wave (CW) spectra on LaF$_3$ to probe the effects of produced $B_1$ on the F--$\mu$--F state populations. A static longitudinal field of $B_0=9.6$~mT was applied, corresponding to a muon Larmor frequency $f_\mu=(\gamma_\mu/2\pi)B_0\approx 1.32$~MHz, and $B_1$ was varied to control the level populations.
The frequency $f_{\mu}$ was chosen to give multiple precessions within the muon time window $\tau_{\mbox{\footnotesize window}}$ (which in our experiment is typically $\approx 10\tau_{\mu}$ where $\tau_\mu=2.2$\,$\mu$s is the mean muon lifetime).

\begin{figure}[t]
	\begin{centering}
		\includegraphics[width=1.0\columnwidth]{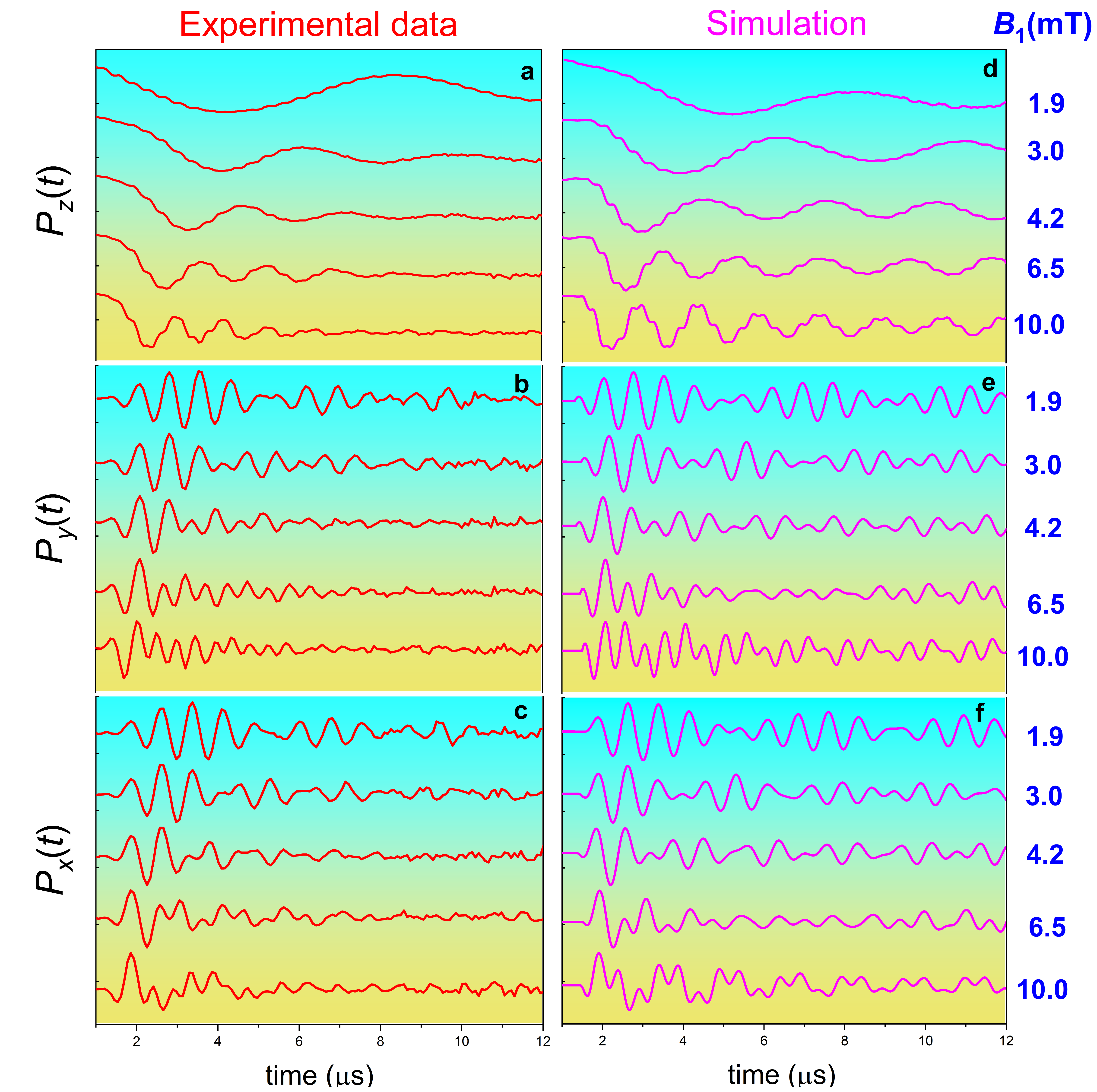} 
		\par\end{centering}
	\caption{The components of both measured and simulated vector polarization $[\vec{P}_x(t)), \vec{P}_y(t), \vec{P}_x(t)]$ under continuous RF excitation for five input power levels parametrised by $B_1$.  The amplitude of the traces have been normalized.}
	\label{fig:B1-modulation}
\end{figure}

As shown in Fig.~\ref{fig:B1-modulation}(a--c), under continuous RF drive in LaF$_3$, the time-resolved muon polarization $P_z(t)$, $P_x(t)$, and $P_y(t)$ shows a clear $B_1$ dependence. As the amplitude of the drive increases (we find $B_1$ is proportional to the square root of the RF power, as shown in Fig.~\ref{fig:Supplementary}), the dominant nutation frequency $\Omega$ rises approximately linearly with $B_1$ ($\Omega \approx \gamma_\mu B_1$ near resonance), seen most cleanly in $P_z(t)$. At higher $B_1$, $P_x(t)$ and $P_y(t)$ develop pronounced oscillatory structures. These structures arise because the measured polarization represents a powder-ensemble average of coherently summed nearby nutation frequencies, whose interference across the orientation distribution generates the observed envelopes.

Small Bloch--Siegert (AC-Zeeman) shifts $\propto B_1^2$ and slight detunings further separate the components \cite{BlochSiegert-1,saiko2008effect}, accentuating the beats, whereas $B_0/B_1$ inhomogeneity primarily increases damping without creating organized nodes. Simulations of the coupled F--$\mu$--F Hamiltonian [Fig.~\ref{fig:B1-modulation}(d--f)], using the same $B_1$--power calibration, reproduce both the monotonic frequency scaling and the emergence of multi-frequency structure in $P_x$ and $P_y$, confirming that the features are an intrinsic consequence of multi-transition driving under CW conditions.  The excellent match between data and simulation confirms that our model captures the driven three‐spin dynamics. Together, these results confirm that we are able to implement genuine coherent Rabi driving in the F–$\mu$–F two‐level manifold, in direct analogy with qubit rotations across leading circuit and spin‐qubit architectures~\cite{pioro2008electrically, nakamura1999coherent}. In other words, by adjusting the duration and amplitude of the RF pulse, this method allows us to realize arbitrary rotation angles (\(\tfrac{\pi}{2}\), \(\pi\), \(\tfrac{3\pi}{2}\), \(\dots\)), effectively implementing universal single‐qubit gates on the Bloch sphere.

\begin{figure}[htbp]
	\begin{centering}
		\includegraphics[width=1\columnwidth]{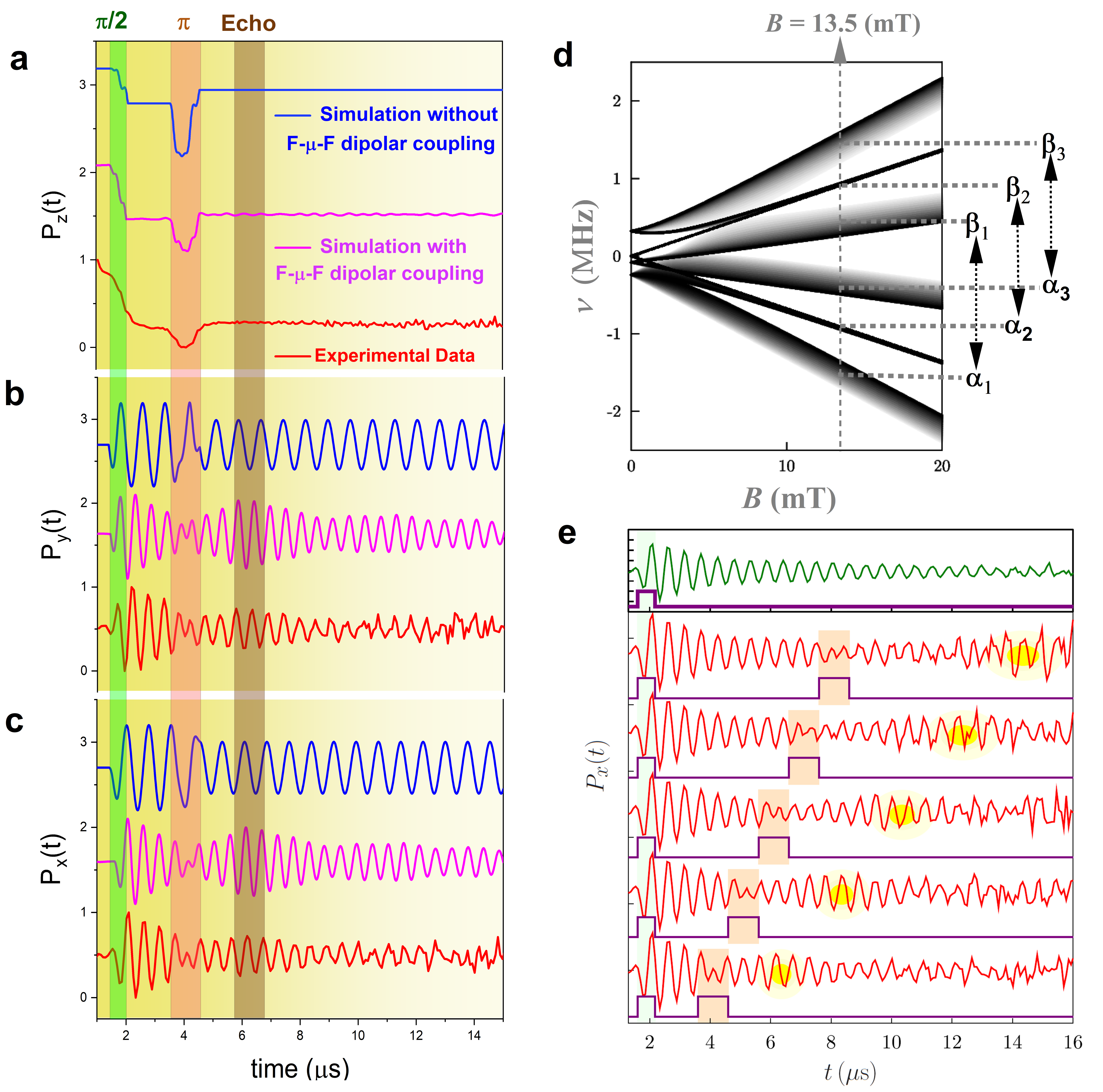} 
		\par\end{centering}
	\caption{ (a--c) The vector polarization $[{P}_z(t)), {P}_y(t), {P}_x(t)]$ following a Hahn-echo sequence (a $\pi/2$ pulse followed by a $\pi$ pulse. The experimental data (red) are shown together with simulations that ignore (blue) and include (pink) the dipolar coupling with the fluorine nuclei. (d) The Breit-Rabi diagram for the F--$\mu$--F state in LaF$_3$ with various transitions highlighted. The grey dotted vertical line at $B=13.5~\mathrm{mT}$ marks the longitudinal field used for the Hahn-echo measurements shown in panels (a--c).(e) Echo position tracks the interpulse delays $\tau$. Red traces show $P_x(t)$ for successive $\tau$; shifting the muon RF $\pi$ pulse (orange) to time $\tau$ after the $\pi/2$ pulse moves the refocused echo (marked in yellow) to $\approx 2\tau$ after the $\pi/2$ pulse.  Purple traces show the pulse sequence that is applied.  The top (green) trace shows the effect of no $\pi$ pulse being applied, so that there is no refocusing. }
	\label{fig:muonecho}
\end{figure}

\section*{Muon spin echo}

In our RF-$\mu$SR-Hahn-echo experiment, a square pulse of duration \(t_p\) delivers a tipping angle $\theta = \gamma_\mu B_1 t_p = \pi/2$ so as to rotate the muon polarization into the transverse plane. The copper-tape  coil delivers excellent \(B_1\) uniformity across the powdered \(\mathrm{LaF}_3\), ensuring consistent flip angles throughout the entire sample. During the free-evolution interval \(\tau\), spins dephase under the frozen dipolar fields of neighboring fluorines with a characteristic decoherence time \(T_2^*\) that far exceeds both the \(\pi/2\) pulse length and the muon time window $\tau_{\mbox{\footnotesize window}}$. % NB tau_window now defined in previous section
At \(t=\tau\), a \(\pi\) (inversion) pulse reverses every accrued phase, so that after a second interval \(\tau\) all spin vectors reconverge into a sharp, amplified echo. Since \(T_2 \gg t_p\) and \(\tau_{\mbox{\footnotesize window}} < T_2\), the echo appears as a robust revival of the time domain signal, providing an enhanced semiclassical fingerprint of dipolar inhomogeneity and residual spin–spin couplings in the \(\mathrm{LaF}_3\) lattice, rather than as a sharp Hahn echo response emerging from a completely decohered background (which is what is normally observed in NMR experiments \cite{Hahn1950}). To precisely model this echo the F--$\mu$--F dipolar interaction is included in our simulations. The muon is not isolated and so does precess at its Larmor frequency $\omega_0 = \gamma_\mu B_0,$ but instead the muon transition frequencies are split into $\omega_0 \pm \omega_{\mathrm{dip}}$ due to the dipolar Hamiltonian \eqref{eq:H_dipole}, which divides the eight‐level F--$\mu$--F manifold.  Each fluorine nucleus contributes a local field $\pm B_{\mathrm{dip}}$ at the muon site, so that the muon “sees” an effective field $B_0 \pm B_{\mathrm{dip}}$. Fig.~\ref{fig:muonecho} \textbf{a--c} show the result of such an experiment (for a $\pi/2$ pulse of length $t_{\pi/2}=0.58$ $\micro$s, $\omega_{\rm RF}/(2\pi)= 1.83 ~ \mathrm{MHz} $ and $B_0=13.5~\mathrm{mT}$).  Inhomogeneities in the dipolar field cause the transverse magnetization to dephase, yet due to the long coherence time of the F--$\mu$--F complex and a nearly homogeneous $B_1$ field, the transverse components $P_x$ and $P_y$ decay slowly. The decaying net transverse magnetization resulting from spin dephasing is refocused by a $\pi$ pulse ($t_{\pi} = 1.02\ \micro\mathrm{s}$) applied after a delay ($\tau = 1.45\ \micro\mathrm{s}$).  The spin‐fan‐out vectors then reconverge at $2\tau$ to form a Hahn echo.  Our simulations (magenta) match the experimental data (red), exhibiting the same oscillation frequencies $\omega_{0} \pm \omega_{\mathrm{dip}}$ and nearly identical decay and echo amplitudes. By contrast, when the dipolar term is omitted from the simulation, the system reduces to a lone spin–$\tfrac{1}{2}$ in $B_{0}$, yielding only a single‐frequency sinusoid at $\omega_{0}$ (blue traces), since there is no local field inhomogeneity to generate additional spectral components and hence no echo effect. This agreement confirms that the F–$\mu$–F dipolar coupling, and the quantum entanglement associated with it, is responsible for the observed change in muon precession frequencies and the rich transverse dynamics in our Hahn-echo measurements.

From a quantum perspective, the Hahn echo RF pulse sequence ($\pi/2$--$\tau$--$\pi$) generates and manipulates coherent superpositions between the Zeeman-split eigenstates of the F--$\mu$--F three-spin complex, selectively driving transitions as illustrated in Fig.~\ref{fig:muonecho}\textbf{d}. The initial $\pi/2$ pulse tips the muon spin into the transverse plane and creates a coherent superposition with amplitude distributed across the $\alpha_i$ and $\beta_i$ branches for $i=1,2,3$ in the coupled F--$\mu$--F manifold.
During the ensuing free-evolution interval $\tau$, the system evolves unitarily under the total Hamiltonian [Eq.~\ref{eq:H_total_time}], and quantum coherence between pairs of eigenstates accumulates phase at rates set by their energy separation (the frequency difference between Zeeman branches), given by
\begin{equation}
  \Delta\phi_{\alpha_i,\beta_i}(t)
  = \big(\lambda_{\alpha_i}-\lambda_{\beta_i}\big) \frac{t}{\hbar}, \qquad i\in\{1,2,3\},
\end{equation}
where $\lambda_{\alpha_i}$ and $\lambda_{\beta_i}$ are the instantaneous eigenvalues of the Hamiltonian [see Eq.~(\ref{eq:diagonalized})]. This phase evolution results in a precession of the muon polarization at characteristic frequencies determined by both dipolar and Zeeman interactions. The subsequent $\pi$ pulse inverts the system, reversing the phase evolution and leading, after a further interval $\tau$, to partial rephasing and the formation of a muon Hahn echo. The detailed structure of the observed signal with each oscillation in the $(P_x,P_y)$-plane time trace mapping onto coherent two-level dynamics governed by the time-dependent Hamiltonian \eqref{eq:H_total_time} thus encodes the coherent quantum dynamics of the F--$\mu$--F qubit.

Varying the delay ($\tau$) between pulses $\tau$ shifts a refocused feature to later times, with its centre following $t_{\mathrm{echo}}(\tau)=2\tau$ while maintaining a modest enhancement of the amplitude relative to the surrounding signal, as shown in Fig~\ref{fig:muonecho}\textbf{e}. This $\tau$-locked translation identifies the feature as a Hahn echo. In contrast, the peaks arising from simple precession or spectral beating remain anchored to the time origin and do not follow $\sim$ 2$\tau$.

\section* {Multi-frequency double resonance}

In experiments with RF pulses in F--$\mu$--F states, an important question is whether coherence can be transferred between the muon and nearby nuclear spins and back again. To achieve this, we need to be able to separately address the muon and fluorine spins using separate pulses tuned to the muon and fluorine spins.  In this double resonance approach \cite{Slitcher, Nu-D}, we insert a fluorine inversion pulse at the midpoint of the Hahn echo sequence, $(\pi/2)_\mu$–$\tau$–$\pi_\text{F}$–$\tau$–echo, where the initial $(\pi/2)_\mu$ pulse uses the same $\omega_{\rm RF}$, $B_1$ and $t_{\pi/2}$ as previously (addressing the muon) while the $\pi_\text{F}$ pulse is instead applied at the fluorine Zeeman frequency $f_{\rm F} = \gamma_\text{F} B_0 / (2\pi)$ with amplitude $B_{1{\rm F}}$ and duration $t_{\pi,\text{F}} = \pi / (\gamma_\text{F} B_{1{\rm F}})$. Then after a further time $\tau$  a muon echo is generated whose amplitude and phase shift are a function of the $\mu$–F dipolar coupling modulated by the fluorine inversion.

\begin{figure}[t]
	\begin{centering}
		\includegraphics[width=1.0\columnwidth]{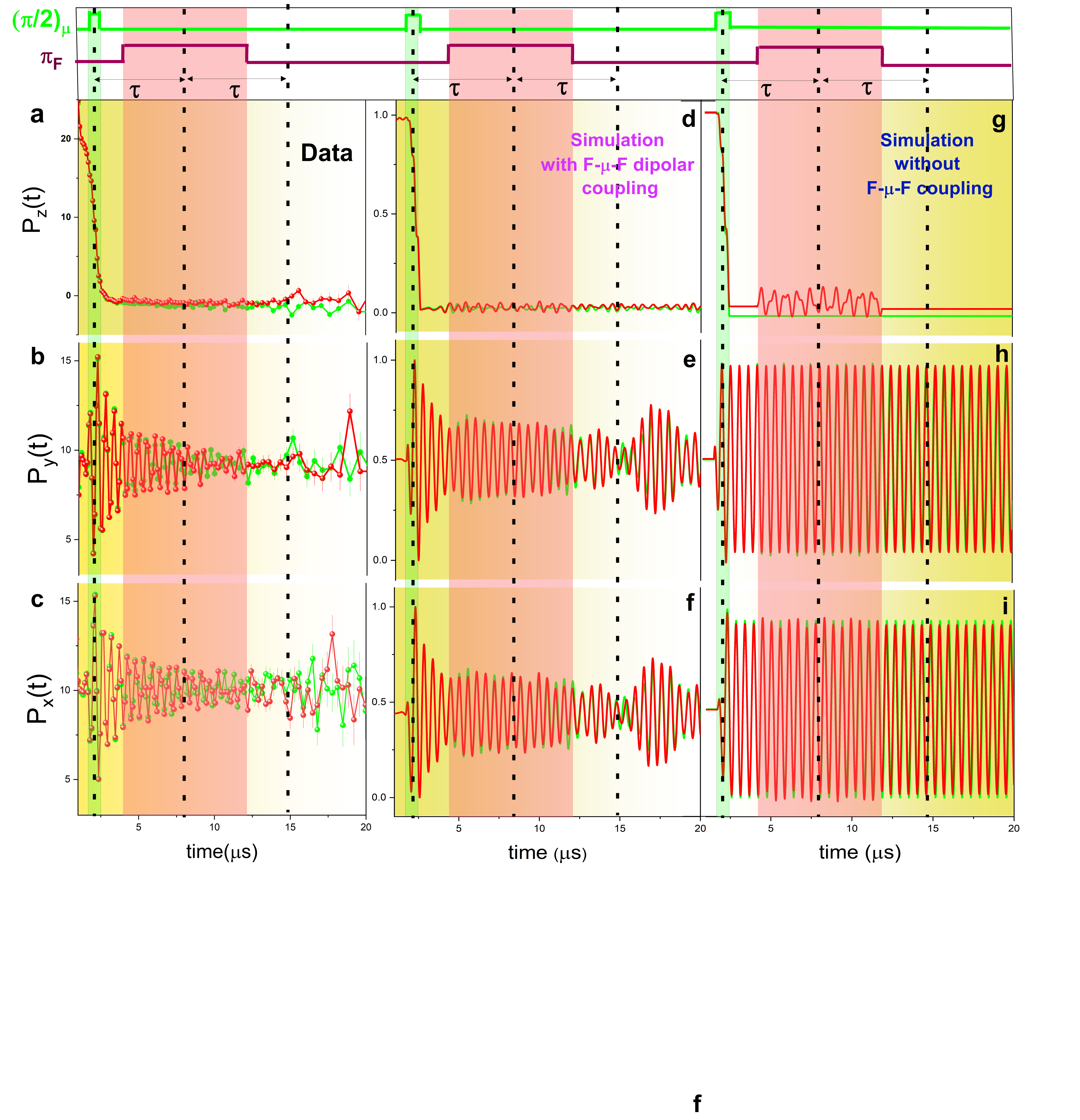} 
		\par\end{centering}
	\caption{(a--c) Vector-resolved muon polarization  $\mathbf P(t)=(P_z,P_y,P_x)$ at the working field of 13.5 mT where $\mu$-only reference [green; single $(\pi/2)_\mu$] versus double resonance [red; $(\pi/2)_\mu$ followed by fluorine $\pi_\text{F}$]. The fluorine $\pi$ pulse produces a discernible amplitude/phase perturbation of the muon signal that becomes pronounced in the region after a time 2$\tau$ has elapsed since the original $(\pi/2)_\mu$ pulse. These plots use variable binning to help resolve the features at late times where the statistics are poor, due to the fact that most muons have decayed at late times. (d--f) Our simulations including the F--$\mu$--F dipolar Hamiltonian reproduce the observed modification of the red trace relative to the green trace (but of course do not suffer from the effect of the finite muon lifetime). 
    (g--i) If the F--$\mu$--F dipolar term is removed from the Hamiltonian in the simulations, the fluorine inversion has no effect and the two traces are indistinguishable, as well as failing to reproduce the main features of the data, confirming the important role of the $\mu$–F coupling in these experiments.}
    \label{fig:DoubleResonance}
\end{figure}

 Fig.~\ref{fig:DoubleResonance}(a--c) shows polarization data recorded for this double resonance pulse sequence: the red trace corresponds to the case where both a $(\pi/2)_\mu$  pulse and a $\pi_\text{F}$ pulse are applied, while the green trace corresponds to just applying the $(\pi/2)_\mu$  pulse. 
 At a time $2\tau$ after the $(\pi/2)_\mu$  pulse, when dephased spins reconverge and the accumulated phases are refocused by the $\pi_\text{F}$ pulse, we observe a small but distinct feature in both the experimental data and simulations that include $\mu$–F dipolar coupling (Fig.~\ref{fig:DoubleResonance}\,(d--f)). 
 However, this echo feature is not very distinct and this is due to several factors. 
 Principally, since $\gamma_{\mbox{F}}$ is approximately one-third of $\gamma_{\mu}$, the $\pi_\text{F}$ pulse has a very long duration, pushing the echo signal to late times $t\gg\tau_\mu$ where the statistics are rather poor. Furthermore, technical constraints on the maximum RF amplitude $B_{1}$ and on pulse shaping further restrict performance.  It should be noted that even in the simulations, the echo is predicted to occur toward the later side of the time window, and the phase accumulation after an ideal $\pi_\text{F}$ pulse does not appear exactly at $2\tau$. To understand these anomalies in the resurgence of phase and amplitude,  note that 
 the $\pi_\text{F}$ pulse inverts the fluorine spins and connects sublevels within each $\mu$-spin family (e.g., $\alpha_1\!\leftrightarrow\!\alpha_2\!\leftrightarrow\!\alpha_3$ and $\beta_1\!\leftrightarrow\!\beta_2\!\leftrightarrow\!\beta_3$).
 Then, after applying a $\pi_\text{F}$ pulse, the system behaves as an effective three‐level system, so that the phases accumulated during free evolution cannot refocus simultaneously across all energy levels. Instead of a clean Hahn echo, one observes an incomplete echo arising from the partial convergence of three-phase trajectories.

\section*{Conclusion}\label{sec13}
The methods described in this paper combine three-dimensional readout of the time dependence of the muon spin state $\vec{P}(t)$ with RF pulse control of individual coherences in a muon--nuclear-spin cluster. The results open up the possibility of deploying more complex pulse sequences. Already, it is clear that this approach revolutionizes the traditional approach of a $\mu$SR experiment in which a muon is implanted in a sample, interacts with local spins, and provides a response that reflects those interactions, but without any further intervention from the experimenter.  Here, the application of targeted pulses to control the muon spin and other spins to which it is coupled, a perturbation which can be easily switched on and off to forensically monitor its effect, opens up new possibilities for studying coupled spin systems.

\bibliography{F-mu-F}


\begin{thebibliography}{21}
\ifx \bisbn   \undefined \def \bisbn  #1{ISBN #1}\fi
\ifx \binits  \undefined \def \binits#1{#1}\fi
\ifx \bauthor  \undefined \def \bauthor#1{#1}\fi
\ifx \batitle  \undefined \def \batitle#1{#1}\fi
\ifx \bjtitle  \undefined \def \bjtitle#1{#1}\fi
\ifx \bvolume  \undefined \def \bvolume#1{\textbf{#1}}\fi
\ifx \byear  \undefined \def \byear#1{#1}\fi
\ifx \bissue  \undefined \def \bissue#1{#1}\fi
\ifx \bfpage  \undefined \def \bfpage#1{#1}\fi
\ifx \blpage  \undefined \def \blpage #1{#1}\fi
\ifx \burl  \undefined \def \burl#1{\textsf{#1}}\fi
\ifx \doiurl  \undefined \def \doiurl#1{\url{https://doi.org/#1}}\fi
\ifx \betal  \undefined \def \betal{\textit{et al.}}\fi
\ifx \binstitute  \undefined \def \binstitute#1{#1}\fi
\ifx \binstitutionaled  \undefined \def \binstitutionaled#1{#1}\fi
\ifx \bctitle  \undefined \def \bctitle#1{#1}\fi
\ifx \beditor  \undefined \def \beditor#1{#1}\fi
\ifx \bpublisher  \undefined \def \bpublisher#1{#1}\fi
\ifx \bbtitle  \undefined \def \bbtitle#1{#1}\fi
\ifx \bedition  \undefined \def \bedition#1{#1}\fi
\ifx \bseriesno  \undefined \def \bseriesno#1{#1}\fi
\ifx \blocation  \undefined \def \blocation#1{#1}\fi
\ifx \bsertitle  \undefined \def \bsertitle#1{#1}\fi
\ifx \bsnm \undefined \def \bsnm#1{#1}\fi
\ifx \bsuffix \undefined \def \bsuffix#1{#1}\fi
\ifx \bparticle \undefined \def \bparticle#1{#1}\fi
\ifx \barticle \undefined \def \barticle#1{#1}\fi
\bibcommenthead
\ifx \bconfdate \undefined \def \bconfdate #1{#1}\fi
\ifx \botherref \undefined \def \botherref #1{#1}\fi
\ifx \url \undefined \def \url#1{\textsf{#1}}\fi
\ifx \bchapter \undefined \def \bchapter#1{#1}\fi
\ifx \bbook \undefined \def \bbook#1{#1}\fi
\ifx \bcomment \undefined \def \bcomment#1{#1}\fi
\ifx \oauthor \undefined \def \oauthor#1{#1}\fi
\ifx \citeauthoryear \undefined \def \citeauthoryear#1{#1}\fi
\ifx \endbibitem  \undefined \def \endbibitem {}\fi
\ifx \bconflocation  \undefined \def \bconflocation#1{#1}\fi
\ifx \arxivurl  \undefined \def \arxivurl#1{\textsf{#1}}\fi
\csname PreBibitemsHook\endcsname

%%% 1
\bibitem[\protect\citeauthoryear{Chuang et~al.}{1998}]{chuang1998experimental}
\begin{barticle}
\bauthor{\bsnm{Chuang}, \binits{I.L.}},
\bauthor{\bsnm{Vandersypen}, \binits{L.M.}},
\bauthor{\bsnm{Zhou}, \binits{X.}},
\bauthor{\bsnm{Leung}, \binits{D.W.}},
\bauthor{\bsnm{Lloyd}, \binits{S.}}:
\batitle{{Experimental realization of a quantum algorithm}}.
\bjtitle{Nature}
\bvolume{393}(\bissue{6681}),
\bfpage{143}--\blpage{146}
(\byear{1998})
\doiurl{10.1038/30181}
\end{barticle}
\endbibitem

%%% 2
\bibitem[\protect\citeauthoryear{Jones}{2000}]{jones2000nmr}
\begin{barticle}
\bauthor{\bsnm{Jones}, \binits{J.}}:
\batitle{{NMR quantum computation: A critical evaluation}}.
\bjtitle{Fortschr. Phys.}
\bvolume{48}(\bissue{9-11}),
\bfpage{909}--\blpage{924}
(\byear{2000})
\doiurl{10.1002/1521-3978(200009)48:9/11<909::AID-PROP909>3.0.CO;2-2}
\end{barticle}
\endbibitem

%%% 3
\bibitem[\protect\citeauthoryear{Knill et~al.}{2000}]{knill2000theory}
\begin{barticle}
\bauthor{\bsnm{Knill}, \binits{E.}},
\bauthor{\bsnm{Laflamme}, \binits{R.}},
\bauthor{\bsnm{Viola}, \binits{L.}}:
\batitle{Theory of quantum error correction for general noise}.
\bjtitle{Phys. Rev. Lett.}
\bvolume{84},
\bfpage{2525}--\blpage{2528}
(\byear{2000})
\doiurl{10.1103/PhysRevLett.84.2525}
\end{barticle}
\endbibitem

%%% 4
\bibitem[\protect\citeauthoryear{Price et~al.}{1999}]{price1999construction}
\begin{barticle}
\bauthor{\bsnm{Price}, \binits{M.}},
\bauthor{\bsnm{Somaroo}, \binits{S.}},
\bauthor{\bsnm{Tseng}, \binits{C.}},
\bauthor{\bsnm{Gore}, \binits{J.}},
\bauthor{\bsnm{Fahmy}, \binits{A.}},
\bauthor{\bsnm{Havel}, \binits{T.}},
\bauthor{\bsnm{Cory}, \binits{D.G.}}:
\batitle{{Construction and implementation of NMR quantum logic gates for two spin systems}}.
\bjtitle{J. Mag. Res.}
\bvolume{140}(\bissue{2}),
\bfpage{371}--\blpage{378}
(\byear{1999})
\doiurl{10.1006/jmre.1999.1851}
\end{barticle}
\endbibitem

%%% 5
\bibitem[\protect\citeauthoryear{Somaroo et~al.}{1999}]{somaroo1999quantum}
\begin{barticle}
\bauthor{\bsnm{Somaroo}, \binits{S.}},
\bauthor{\bsnm{Tseng}, \binits{C.H.}},
\bauthor{\bsnm{Havel}, \binits{T.F.}},
\bauthor{\bsnm{Laflamme}, \binits{R.}},
\bauthor{\bsnm{Cory}, \binits{D.G.}}:
\batitle{Quantum simulations on a quantum computer}.
\bjtitle{Phys. Rev. Lett.}
\bvolume{82},
\bfpage{5381}--\blpage{5384}
(\byear{1999})
\doiurl{10.1103/PhysRevLett.82.5381}
\end{barticle}
\endbibitem

%%% 6
\bibitem[\protect\citeauthoryear{Leskowitz et~al.}{2003}]{leskowitz2003three}
\begin{barticle}
\bauthor{\bsnm{Leskowitz}, \binits{G.M.}},
\bauthor{\bsnm{Ghaderi}, \binits{N.}},
\bauthor{\bsnm{Olsen}, \binits{R.A.}},
\bauthor{\bsnm{Mueller}, \binits{L.J.}}:
\batitle{{Three-qubit nuclear magnetic resonance quantum information processing with a single-crystal solid}}.
\bjtitle{J. Chem. Phys.}
\bvolume{119}(\bissue{3}),
\bfpage{1643}--\blpage{1649}
(\byear{2003})
\doiurl{10.1063/1.1582171}
\end{barticle}
\endbibitem

%%% 7
\bibitem[\protect\citeauthoryear{Slichter}{2013}]{Slitcher}
\begin{bbook}
\bauthor{\bsnm{Slichter}, \binits{C.P.}}:
\bbtitle{Principles of Magnetic Resonance},
\bedition{3}rd edn.
\bpublisher{Springer},
\blocation{Berlin, Heidelberg}
(\byear{2013}).
\doiurl{10.1007/978-3-662-09441-9}
\end{bbook}
\endbibitem

%%% 8
\bibitem[\protect\citeauthoryear{Levitt}{2008}]{Levitt}
\begin{bbook}
\bauthor{\bsnm{Levitt}, \binits{M.H.}}:
\bbtitle{Spin Dynamics},
\bedition{2nd} edn.
\bpublisher{John Wiley \& Sons},
\blocation{Chichester, England}
(\byear{2008})
\end{bbook}
\endbibitem

%%% 9
\bibitem[\protect\citeauthoryear{Clayden et~al.}{2012}]{Clayden2012}
\begin{barticle}
\bauthor{\bsnm{Clayden}, \binits{N.J.}},
\bauthor{\bsnm{Cottrell}, \binits{S.P.}},
\bauthor{\bsnm{McKenzie}, \binits{I.}}:
\batitle{Spin evolution in a radio frequency field studied through muon spin resonance}.
\bjtitle{Journal of Magnetic Resonance}
\bvolume{214},
\bfpage{144}--\blpage{150}
(\byear{2012})
\doiurl{10.1016/j.jmr.2011.10.018}
\end{barticle}
\endbibitem

%%% 10
\bibitem[\protect\citeauthoryear{Brewer et~al.}{1986}]{PhysRevBBrewerF-mu-F}
\begin{barticle}
\bauthor{\bsnm{Brewer}, \binits{J.H.}},
\bauthor{\bsnm{Kreitzman}, \binits{S.R.}},
\bauthor{\bsnm{Noakes}, \binits{D.R.}},
\bauthor{\bsnm{Ansaldo}, \binits{E.J.}},
\bauthor{\bsnm{Harshman}, \binits{D.R.}},
\bauthor{\bsnm{Keitel}, \binits{R.}}:
\batitle{{Observation of muon-fluorine "hydrogen bonding" in ionic crystals}}.
\bjtitle{Phys. Rev. B}
\bvolume{33},
\bfpage{7813}--\blpage{7816}
(\byear{1986})
\doiurl{10.1103/PhysRevB.33.7813}
\end{barticle}
\endbibitem

%%% 11
\bibitem[\protect\citeauthoryear{Lancaster et~al.}{2007}]{PhysRevLett.99.267601}
\begin{barticle}
\bauthor{\bsnm{Lancaster}, \binits{T.}},
\bauthor{\bsnm{Blundell}, \binits{S.J.}},
\bauthor{\bsnm{Baker}, \binits{P.J.}},
\bauthor{\bsnm{Brooks}, \binits{M.L.}},
\bauthor{\bsnm{Hayes}, \binits{W.}},
\bauthor{\bsnm{Pratt}, \binits{F.L.}},
\bauthor{\bsnm{Manson}, \binits{J.L.}},
\bauthor{\bsnm{Conner}, \binits{M.M.}},
\bauthor{\bsnm{Schlueter}, \binits{J.A.}}:
\batitle{{Muon-Fluorine Entangled States in Molecular Magnets}}.
\bjtitle{Phys. Rev. Lett.}
\bvolume{99},
\bfpage{267601}
(\byear{2007})
\doiurl{10.1103/PhysRevLett.99.267601}
\end{barticle}
\endbibitem

%%% 12
\bibitem[\protect\citeauthoryear{Wilkinson and Blundell}{2020}]{PhysRevLett.125.087201}
\begin{barticle}
\bauthor{\bsnm{Wilkinson}, \binits{J.M.}},
\bauthor{\bsnm{Blundell}, \binits{S.J.}}:
\batitle{{Information and Decoherence in a Muon-Fluorine Coupled System}}.
\bjtitle{Phys. Rev. Lett.}
\bvolume{125},
\bfpage{087201}
(\byear{2020})
\doiurl{10.1103/PhysRevLett.125.087201}
\end{barticle}
\endbibitem

%%% 13
\bibitem[\protect\citeauthoryear{Kreitzman et~al.}{1988}]{PhysRevLett.61.2890}
\begin{barticle}
\bauthor{\bsnm{Kreitzman}, \binits{S.R.}},
\bauthor{\bsnm{Williams}, \binits{D.L.}},
\bauthor{\bsnm{Kaplan}, \binits{N.}},
\bauthor{\bsnm{Kempton}, \binits{J.R.}},
\bauthor{\bsnm{Brewer}, \binits{J.H.}}:
\batitle{{Spin Echoes for ${\ensuremath{\mu}}^{+}$-Spin Spectroscopy}}.
\bjtitle{Phys. Rev. Lett.}
\bvolume{61},
\bfpage{2890}--\blpage{2893}
(\byear{1988})
\doiurl{10.1103/PhysRevLett.61.2890}
\end{barticle}
\endbibitem

%%% 14
\bibitem[\protect\citeauthoryear{Billington et~al.}{2022}]{PhysRevLettBillington}
\begin{barticle}
\bauthor{\bsnm{Billington}, \binits{D.}},
\bauthor{\bsnm{Riordan}, \binits{E.}},
\bauthor{\bsnm{Salman}, \binits{M.}},
\bauthor{\bsnm{Margineda}, \binits{D.}},
\bauthor{\bsnm{Gill}, \binits{G.J.W.}},
\bauthor{\bsnm{Cottrell}, \binits{S.P.}},
\bauthor{\bsnm{McKenzie}, \binits{I.}},
\bauthor{\bsnm{Lancaster}, \binits{T.}},
\bauthor{\bsnm{Graf}, \binits{M.J.}},
\bauthor{\bsnm{Giblin}, \binits{S.R.}}:
\batitle{{Radio-Frequency Manipulation of State Populations in an Entangled Fluorine-Muon-Fluorine System}}.
\bjtitle{Phys. Rev. Lett.}
\bvolume{129},
\bfpage{077201}
(\byear{2022})
\doiurl{10.1103/PhysRevLett.129.077201}
\end{barticle}
\endbibitem

%%% 15
\bibitem[\protect\citeauthoryear{Hahn}{1950}]{Hahn1950}
\begin{barticle}
\bauthor{\bsnm{Hahn}, \binits{E.L.}}:
\batitle{Spin echoes}.
\bjtitle{Phys. Rev.}
\bvolume{80},
\bfpage{580}--\blpage{594}
(\byear{1950})
\doiurl{10.1103/PhysRev.80.580}
\end{barticle}
\endbibitem

%%% 16
\bibitem[\protect\citeauthoryear{Yan et~al.}{2015}]{BlochSiegert-1}
\begin{barticle}
\bauthor{\bsnm{Yan}, \binits{Y.}},
\bauthor{\bsnm{L\"u}, \binits{Z.}},
\bauthor{\bsnm{Zheng}, \binits{H.}}:
\batitle{{Bloch-Siegert shift of the Rabi model}}.
\bjtitle{Phys. Rev. A}
\bvolume{91},
\bfpage{053834}
(\byear{2015})
\doiurl{10.1103/PhysRevA.91.053834}
\end{barticle}
\endbibitem

%%% 17
\bibitem[\protect\citeauthoryear{Saiko and Fedoruk}{2008}]{saiko2008effect}
\begin{barticle}
\bauthor{\bsnm{Saiko}, \binits{A.P.}},
\bauthor{\bsnm{Fedoruk}, \binits{G.}}:
\batitle{{Effect of the Bloch-Siegert shift on the frequency responses of Rabi oscillations in the case of nutation resonance}}.
\bjtitle{JETP Lett.}
\bvolume{87}(\bissue{3}),
\bfpage{128}--\blpage{132}
(\byear{2008})
\doiurl{10.1134/S002136400803003X}
\end{barticle}
\endbibitem

%%% 18
\bibitem[\protect\citeauthoryear{Pioro-Ladriere et~al.}{2008}]{pioro2008electrically}
\begin{barticle}
\bauthor{\bsnm{Pioro-Ladriere}, \binits{M.}},
\bauthor{\bsnm{Obata}, \binits{T.}},
\bauthor{\bsnm{Tokura}, \binits{Y.}},
\bauthor{\bsnm{Shin}, \binits{Y.-S.}},
\bauthor{\bsnm{Kubo}, \binits{T.}},
\bauthor{\bsnm{Yoshida}, \binits{K.}},
\bauthor{\bsnm{Taniyama}, \binits{T.}},
\bauthor{\bsnm{Tarucha}, \binits{S.}}:
\batitle{{Electrically driven single-electron spin resonance in a slanting Zeeman field}}.
\bjtitle{Nature Physics}
\bvolume{4}(\bissue{10}),
\bfpage{776}--\blpage{779}
(\byear{2008})
\doiurl{10.1038/nphys1053}
\end{barticle}
\endbibitem

%%% 19
\bibitem[\protect\citeauthoryear{Nakamura et~al.}{1999}]{nakamura1999coherent}
\begin{barticle}
\bauthor{\bsnm{Nakamura}, \binits{Y.}},
\bauthor{\bsnm{Pashkin}, \binits{Y.A.}},
\bauthor{\bsnm{Tsai}, \binits{J.}}:
\batitle{{Coherent control of macroscopic quantum states in a single-Cooper-pair box}}.
\bjtitle{Nature}
\bvolume{398}(\bissue{6730}),
\bfpage{786}--\blpage{788}
(\byear{1999})
\doiurl{10.1038/19718}
\end{barticle}
\endbibitem

%%% 20
\bibitem[\protect\citeauthoryear{Hartmann and Hahn}{1962}]{Nu-D}
\begin{barticle}
\bauthor{\bsnm{Hartmann}, \binits{S.R.}},
\bauthor{\bsnm{Hahn}, \binits{E.L.}}:
\batitle{Nuclear double resonance in the rotating frame}.
\bjtitle{Phys. Rev.}
\bvolume{128},
\bfpage{2042}--\blpage{2053}
(\byear{1962})
\doiurl{10.1103/PhysRev.128.2042}
\end{barticle}
\endbibitem

%%% 21
\bibitem[\protect\citeauthoryear{Cottrell et~al.}{1997}]{Cottrell1997}
\begin{barticle}
\bauthor{\bsnm{Cottrell}, \binits{S.P.}},
\bauthor{\bsnm{Scott}, \binits{C.A.}},
\bauthor{\bsnm{Hitti}, \binits{B.}}:
\batitle{The development of a facility for radio‐frequency experiments at isis}.
\bjtitle{Hyp. Int.}
\bvolume{106},
\bfpage{251}--\blpage{256}
(\byear{1997})
\doiurl{10.1023/A:1012666613777}
\end{barticle}
\endbibitem

\end{thebibliography}

\backmatter

\small

\section*{Methods}
\subsection* {Simulation Method}\label{sec2}
 Each of the three spins $\large(S = \tfrac12\large)$ couples to the others through magnetic dipole--dipole interactions, experiences Zeeman splitting in a static external field $B_0$, and can be driven by a transverse oscillating RF field $B_1(t)$; together, these terms govern the time-dependent evolution of the entire spin ensemble. To capture this behavior, we represent the system in an eight-dimensional Hilbert space (the tensor product of three two-level spins) and integrate the Liouville--von Neumann equation under the full time-dependent Hamiltonian over fine time steps. This framework produces simulations of $\mu^+$-spin polarization dynamics for direct comparison with RF-$\mu$SR measurements.
\vspace{1ex}

\noindent\textbf{Physical constants and parameters:}  The gyromagnetic ratio of the muon is $\gamma_{\mu}/2\pi = 135.53~\mathrm{MHz/T}$ and that of fluorine is $\gamma_\text{F}/2\pi = 40.053~\mathrm{MHz/T}$. A static field $B_{0}$ establishes the Zeeman splitting of the entangled F–$\mu$–F complex; the system is then driven in sequence by two square RF excitations at frequencies $f_{1}$ and $f_{2}$ with amplitudes $B_{1}$ and $B_{1,2}$ over the intervals $[t_{\mathrm{start1}},\,t_{\mathrm{end1}}]$ and $[t_{\mathrm{start2}},\,t_{\mathrm{end2}}]$, respectively.

\vspace{1ex}
\noindent\textbf{Spin Operators and Hilbert Space:}  
The implanted muon and two fluorine nuclei are represented by the Pauli matrices $\sigma_x$, $\sigma_y$, and $\sigma_z$, along with the $2\times2$ identity $I_2$. Each spin-$\tfrac12$ particle occupies a two-dimensional Hilbert space, so the combined system  

\begin{equation}\label{eq:H}
  \mathcal{H}
  = \mathcal{H}_\mu \otimes \mathcal{H}_{\text{F}_1} \otimes \mathcal{H}_{\text{F}_2}
\end{equation}
is eight-dimensional:
\begin{equation}\label{eq:dim}
  \dim\mathcal{H}
  = 2_\mu \times 2_{\text{F}_1} \times 2_{\text{F}_2}
  = 8.
\end{equation}
Operators on $\mathcal{H}$ are constructed using Kronecker products of single-spin operators with identity matrices on the spectator spins. For each $\alpha\in\{x,y,z\}$:
\begin{align}\label{eq:spin_ops}
  \hat S_\alpha
  &= \bigl(\tfrac12\,\sigma_\alpha\bigr)_\mu \otimes I_{\text{F}_1} \otimes I_{\text{F}_2},\\
  \hat I_{\text{F}_1,\alpha}
  &= I_\mu \otimes \bigl(\tfrac12\,\sigma_\alpha\bigr)_{\text{F}_1} \otimes I_{\text{F}_2},\nonumber\\
  \hat I_{\text{F}_2,\alpha}
  &= I_\mu \otimes I_{\text{F}_1} \otimes \bigl(\tfrac12\,\sigma_\alpha\bigr)_{\text{F}_2}.\nonumber
\end{align}
The set $\{\hat S_\alpha,\hat I_{\text{F}_1,\alpha},\hat I_{\text{F}_2,\alpha}\}$ (and $I_{\mathcal{H}}$) form a complete basis for the Hamiltonian, the density matrix, and other observables.
This approach can be extended to include additional fluorine spins \cite{PhysRevLett.125.087201} but we found that our minimal model, which concentrates on the most important couplings between the muon and its two nearest-neighbour fluorine spins, was sufficient to model the experimental data.

\vspace{1ex}
\noindent\textbf{Dipolar Hamiltonian:}  
The magnetic dipole coupling between any two spins $i$ and $j$ separated by a vector $\mathbf{r}_{ij}$ is given by
\begin{equation}\label{eq:H_dipole}
  \hat{H}_{\mathrm{dipole}}^{(ij)} = \frac{\mu_0}{4\pi}\,\frac{\gamma_i\,\gamma_j}{|\mathbf{r}_{ij}|^3}
  \left[\hat{\mathbf{J}}_i\cdot\hat{\mathbf{J}}_j - 3\,(\hat{\mathbf{J}}_i\cdot\hat{\mathbf{r}}_{ij})(\hat{\mathbf{J}}_j\cdot\hat{\mathbf{r}}_{ij})\right],
\end{equation}
where $\hat{\mathbf{J}}_i = (\hat{J}_{i,x}, \hat{J}_{i,y}, \hat{J}_{i,z})$ are the spin-$\tfrac12$ operators and $\hat{\mathbf{r}}_{ij} = \mathbf{r}_{ij} / |\mathbf{r}_{ij}|$. In LaF$_3$, the equilibrium $\mu$–F bond length is $r_{\mu F} \approx 1.20$\,\AA, with the two fluorine nuclei positioned nearly collinear on opposite sides of the muon. Numerically, one fixes the muon at the origin and places the fluorines at $(0, 0, \pm r)$. The total dipolar Hamiltonian of the three spin system is then
\begin{equation}\label{eq:H_ditotal}
  \hat{H}_{\mathrm{dipole}} = \hat{H}_{\mathrm{dipole}}^{(\mu \text{F}_1)} + \hat{H}_{\mathrm{dipole}}^{(\mu \text{F}_2)} + \hat{H}_{\mathrm{dipole}}^{(\text{F}_1 \text{F}_2)},
\end{equation}
with each pair term constructed by substituting the appropriate $\hat{\mathbf{J}}_i$, gyromagnetic ratios, and $\mathbf{r}_{ij}$. To capture arbitrary sample orientations, we take the bond-axis unit vector in spherical coordinates:
\begin{equation}\label{eq:rhat}
  \hat{\mathbf{r}}_{ij} = (\sin\theta_i\cos\phi_i,\, \sin\theta_i\sin\phi_i,\, \cos\theta_i),
\end{equation}
where $\theta_i$ (polar) and $\phi_i$ (azimuthal) are measured from the muon spin quantization ($z$) axis. In the computational implementation, we evaluate scalar products $\hat{\mathbf{J}}_i \cdot \hat{\mathbf{J}}_j$ and projections $\hat{\mathbf{J}}_i \cdot \hat{\mathbf{r}}_{ij} = r_{ij,x} \hat{J}_{i,x} + r_{ij,y} \hat{J}_{i,y} + r_{ij,z} \hat{J}_{i,z}$, each of which becomes an $8\times8$ matrix once the single-spin operators are embedded in the full Hilbert space.

\vspace{1ex}
\noindent\textbf{Zeeman Hamiltonian with pulsed RF:} 
The static external field $B_0$ is taken along the laboratory $z$-axis, so that each spin $i$ experiences a Zeeman interaction
\begin{equation}\label{eq:zeeman_static}
  \hat{H}_{\mathrm{Zeeman},\,stat}^{(i)} = -\gamma_i\,\hat{J}_{i,z}\,B_0.
\end{equation}
To this static configuration we add an oscillating RF field is applied along the laboratory $x$ or $y$ direction (here taken along $x$ in the rotating-wave picture). During the first pulse interval $t_{\mathrm{start1}} \le t \le t_{\mathrm{end1}}$, the field takes the form
\begin{equation}\label{eq:B1}
  \mathbf{B}_1(t) = \bigl( B_1\,\cos(2\pi f_1\,t),\,0,\,0 \bigr),
\end{equation}
which acts on both the muon and the fluorine spins, contributing a time-dependent term
\begin{equation}\label{eq:H_RF1}
  \hat{H}_{\mathrm{RF}}^{(i)}(t) = -\gamma_i\,\hat{J}_{i,x}\,B_1\,\cos(2\pi f_1\,t).
\end{equation}
Similarly, during the second pulse interval $t_{\mathrm{start2}} \le t \le t_{\mathrm{end2}}$, an on-resonance field
\begin{equation}\label{eq:B12}
  \mathbf{B}_{12}(t) = \bigl( B_{12}\,\cos(2\pi f_2\,t),\,0,\,0 \bigr)
\end{equation}
is applied. Outside these pulse windows, the RF field vanishes. The total RF Hamiltonian, summed over all three spins, is
\begin{equation}\label{eq:H_RF_total}
  \hat{H}_{\mathrm{RF}}(t) 
  = -\sum_{i \in \{\mu, \text{F}_1, \text{F}_2\}} \gamma_i\,\hat{J}_{i,x}\,B_{\mathrm{pulse}}(t)\,\cos(2\pi f\,t),
\end{equation}
where $B_{\mathrm{pulse}}(t)$ and $f$ switch between $\{B_1, f_1\}$ and $\{B_{12}, f_2\}$ depending on the time interval.

\vspace{1ex}
\noindent\textbf{Total Hamiltonian and diagonalization:} 
At each discrete time point $t_k = k\,\Delta t$, the total Hamiltonian is assembled as
\begin{equation}\label{eq:H_total_time}
  \hat{H}(t_k) 
  = \hat{H}_{\mathrm{dipole}} + \hat{H}_{\mathrm{Zeeman},\,stat} + \hat{H}_{\mathrm{RF}}(t_k).
\end{equation}
Since $\hat{H}_{\mathrm{dipole}}$ and $\hat{H}_{\mathrm{Zeeman},\,0}$ are time-independent, they can be precomputed once as fixed $8\times8$ matrices. Only the RF term must be updated at each $t_k$. We then diagonalize $\hat{H}(t_k)$ numerically, obtaining the eigenvalues $\Bigr\{\lambda^{(k)}_\alpha\Bigr\}$ and the eigenvectors $\Bigr\{\lvert \phi^{(k)}_\alpha\rangle\Bigr\}$. Denoting by $V^{(k)}$ the unitary matrix whose columns are the eigenvectors, we write
\begin{equation}\label{eq:diagonalized}
  V^{(k)\,\dagger}\,\hat{H}(t_k)\,V^{(k)} = \operatorname{diag}\bigl(\lambda^{(k)}_1,\,\lambda^{(k)}_2,\,\ldots,\,\lambda^{(k)}_8\bigr).
\end{equation}
The short-time propagator over $\Delta t$ is then
\begin{equation}\label{eq:prop}
  U^{(k)} = V^{(k)}\,\exp\!\Bigl(-i\,\Delta t\,\operatorname{diag}\left(\lambda^{(k)}_\alpha\right)\Bigr)\,V^{(k)\,\dagger},
\end{equation}
which advances the density matrix from $\rho(t_k)$ to
\begin{equation}\label{eq:density matrix}
  \rho(t_{k+1}) = U^{(k)}\,\rho(t_k)\,U^{(k)\,\dagger}.
\end{equation}
The use of instantaneous diagonalization at each $t_k$ ensures that fast oscillations from the RF field are accurately resolved on the chosen time grid.

\vspace{1ex}
\noindent\textbf{Initial Density Matrix:}  
At $t = 0$, the muon spin is initialized in a pure state fully polarized along the $+z$ direction, with $\rho_\mu(0) = \lvert\uparrow\rangle\langle\uparrow\rvert = \begin{pmatrix}1 & 0 \\ 0 & 0\end{pmatrix}$. In contrast, each fluorine nucleus is treated as a completely mixed state,
\begin{equation}\label{eq:rho_F}
  \rho_{\text{F}_i}(0) = \tfrac12\,I_2 = \tfrac12 \begin{pmatrix}1 & 0 \\ 0 & 1\end{pmatrix}, \quad i = 1,2.
\end{equation}
Hence the full three-spin initial state is the tensor product
\begin{equation}\label{eq:rho_0}
  \rho(0) = \rho_\mu(0) \otimes \rho_{\text{F}_1}(0) \otimes \rho_{\text{F}_2}(0),
\end{equation}
an $8\times8$ matrix. At each step, we update $\rho(t)$ using the propagator $U^{(k)}$.

\vspace{1ex}
\noindent\textbf{Extraction of Muon Polarization:} The observable of interest is the muon polarization vector
\begin{equation}\label{eq:P_mu_vector}
  \mathbf{P}_\mu(t)
  = \mathrm{Tr}\bigl[\rho(t)\,\hat{\mathbf{S}}\bigr]\times2
  = \bigl(P_\mu^x(t),\,P_\mu^y(t),\,P_\mu^z(t)\bigr),
\end{equation}
where
\[
  \hat{\mathbf{S}} = (\hat S_x,\;\hat S_y,\;\hat S_z)
  \,,\quad
  \hat S_i = \tfrac12\,\sigma_i
  \quad(i=x,y,z)
\]
are the muon’s spin-\(\tfrac12\) operators tensor-expanded to dimension 8.  
The factor of 2 normalizes each expectation value
\(\langle \hat S_i\rangle\in[-\tfrac12,+\tfrac12]\)
to a polarization component
\[
  P_\mu^i(t) \;=\; 2\,\mathrm{Tr}\bigl[\rho(t)\,\hat S_i\bigr]
  \;\in\;[-1,+1]
  \quad(i=x,y,z).
\]
In particular, one may extract not only the longitudinal polarization \(P_\mu^z\) but also the transverse components \(P_\mu^x\) and \(P_\mu^y\).  Numerically, after computing \(\rho(t_k)\) at each time step \(t_k\), one forms
\[
  \rho(t_k)\,\hat S_i,\quad i=x,y,z,
\]
takes the trace, and records
\begin{equation}\label{eq:P_mu_components}
  P_\mu^i(t_k)
  = 2\,\mathrm{Tr}\bigl[\rho(t_k)\,\hat S_i\bigr]
  \quad(i=x,y,z).
\end{equation}

\vspace{1ex}
\noindent\textbf{Orientational Averaging:}  
Since the experiment was not performed on a single crystal, the F--$\mu$--F axis can assume arbitrary orientations relative to the laboratory frame. To account for this, we sample $N$ orientations $\{(\theta_i, \phi_i)\}$ on the unit sphere using a golden-angle prescription to ensure near-uniform coverage. These orientations are used in the construction of the unit vector $\hat{\mathbf{r}}_{ij}$ as defined in Eq.~\eqref{eq:rhat}. For each $(\theta_i, \phi_i)$, we rotate the dipole-coupling quantization axis equivalently, we transform the single-spin operators $\hat{J}_{i,\alpha}$ — such that the internuclear axis aligns with $(\theta_i, \phi_i)$.
 We then perform the full time evolution: Hamiltonian assembly, diagonalization, density-matrix propagation, and muon-polarization extraction, resulting in a polarization trajectory $P_\mu^{(i)}(t_k)$. The orientation-averaged polarization is given by
\begin{equation}\label{eq:orient_avg}
  P_\mu^{\mathrm{avg}}(t_k) = \frac{1}{N} \sum_{i=1}^N P_\mu^{(i)}(t_k),
\end{equation}
which is simply the mean of the $N$ computed trajectories.

\vspace{1ex}

\subsection* {Muon experiment}\label{sec3}
We performed the experiment on the EMU spectrometer at ISIS. Approximately 200 mg of powdered $\mathrm{LaF}_3$ was wrapped in Kapton tape, and mounted in a low-background plastic sample holder at room temperature. A resonant coil was made by tightly winding a $0.1$ mm thick copper tape directly around the sample; this coil was tuned to the F–$\mu$–F transition frequency ($\omega_0/2\pi$) with its quality factor $Q$ adjusted appropriately according to the bandwidth of the resonant circuit used in the experiment. Power-regulated square RF pulsing was generated by a signal generator synchronized to the ISIS accelerator trigger so that each muon pulse arrived at a well-defined point in the sequence.  The beam spot has a diameter of approximately 15 mm, and so the coil has dimensions which are about 20 mm $\times$ 2mm, much larger than a typical NMR coil, which makes achieving high $B_1$ values more challenging.

For the CW experiment, a long-pulse excitation was applied approximately 1.5~$\mu$s after the implantation of the muons into the sample. Because the incoming muons arrive within a pulse of width $\approx 50$~ns, this short delay ensures that the muon pulse has arrived before RF excitation begins so that all muons precess coherently, independent of their precise arrival time. The RF phase is thus locked to the muons which allows the transverse ($P_x$ and $P_y$) polarization to be investigated (see e.g. \cite{Cottrell1997}).

The pulse lengths in the Hahn echo and double resonance experiments were calibrated by measuring $P_z$ as a function of time after applying pulses of different lengths.  The finite rise time at the leading edge of our square pulses is the origin of the fact that $t_\pi$ is not precisely double that of $t_{\pi/2}$.

For the double resonance experiment, the $\pi$ pulse has duration $t_{\pi,\mathrm{fluorine}} =6.9\ \micro\mathrm{s}$, the $\pi/2$ pulse has duration $t_{\pi/2} = 0.84\ \micro\mathrm{s}$, and the inter-pulse delay is $\tau =6.03\ \micro\mathrm{s}$. %All now quoted to 2 decimal places

\setcounter{figure}{0}
\renewcommand{\figurename}{Fig.}
\renewcommand{\thefigure}{S\arabic{figure}}

\section*{Supplementary information}
\begin{figure}[t]
	\begin{centering}
		\includegraphics[width=1.0\columnwidth]{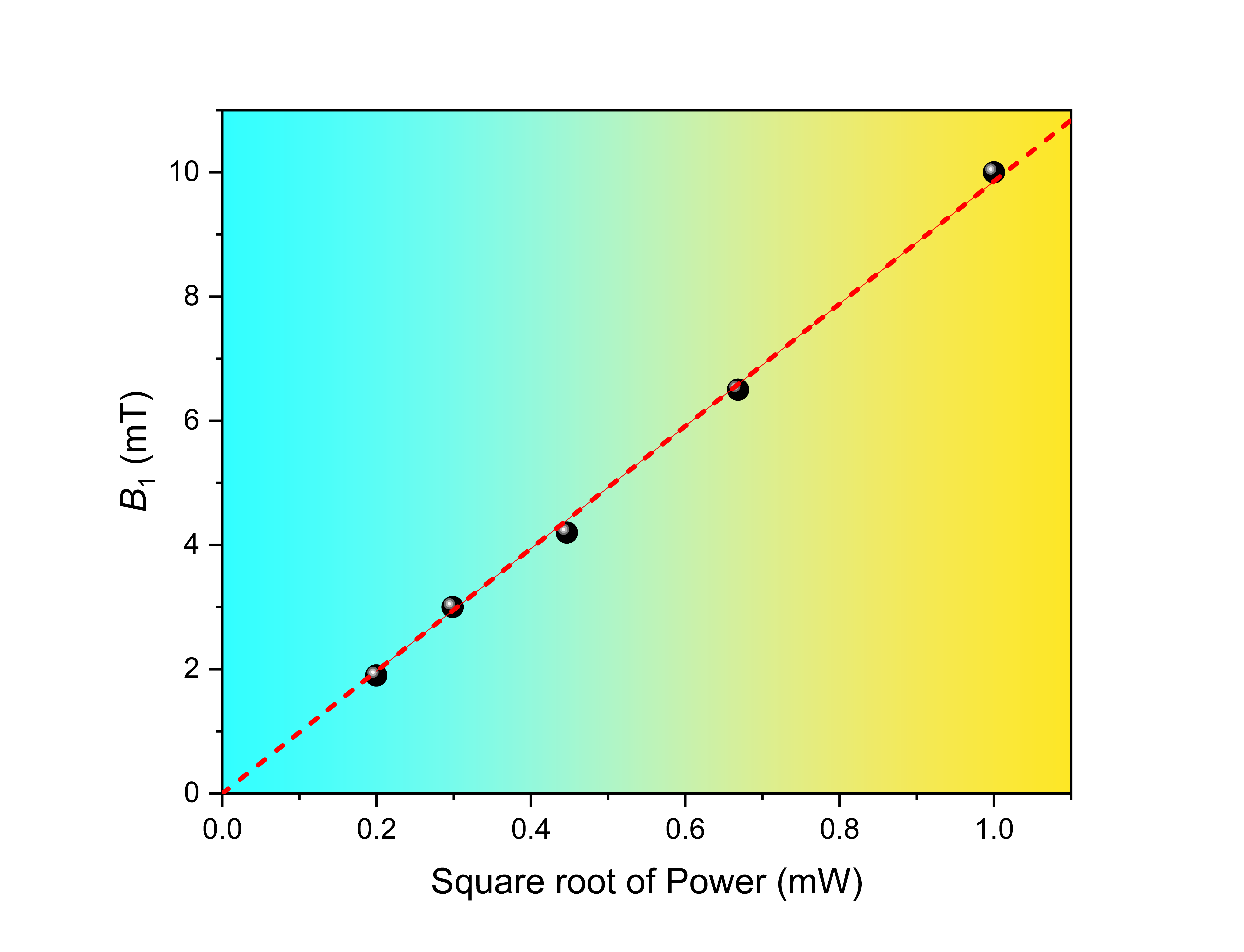} 
		\par\end{centering}
	\caption{Linear dependence of $B_{1}$ on the square root of applied RF power delivered to the power amplifier.}
     \label{fig:Supplementary}
\end{figure}
The applied RF power refers to the power delivered to the power amplifier, which in turn drives the resonator and generates the oscillating magnetic field $B_{1}$.
Figure~\ref{fig:Supplementary} shows that $B_1$ is proportional to the square root of the RF power delivered to the power amplifier.

\section*{Declarations}

\begin{itemize}
\item {\bf Data availability} 

The data that support the findings of
this article are available via the
STFC ISIS Neutron and Muon Source,
10.5286/ISIS.E.RB2510545 (2025).

\item {\bf Acknowledgments}

We thank A. Ardavan and J. M. Wilkinson for useful discussions.
We thank the ERC for the award of an advanced grant and 
acknowledge support by the ISIS Neutron and Muon
Source and the UK Research and Innovation (UKRI)
under the UK government’s Horizon Europe funding
guarantee [Grant No. EP/X025861/1]. 
\item {\bf Author contributions}

SJB conceived the project.  DC, SPC and SJB designed the experiment, which was carried out by DC, BMH, HCHW, SPC and SJB on a sample grown by DP.  The RF-$\mu$SR apparatus has been developed by AL and SPC.  DC and SJB analysed the data and wrote the simulation software.  DC and SJB wrote the paper with input from all authors.

\item {\bf Competing interests}

The authors declare no competing interests.
\end{itemize}

\end{document}